\documentclass{ifacconf}

\usepackage{amsmath,amsfonts,amssymb,graphicx,stmaryrd}
\usepackage{graphics}
\usepackage{natbib}        

\def\tt{{(22)}}

\newcommand{\bd}{\begin{displaymath}}
\newcommand{\ed}{\end{displaymath}}
\newcommand{\bea}{\begin{eqnarray}}
\newcommand{\eea}{\end{eqnarray}}
\newcommand{\Hi}{{\cal H}_\infty}

\newcommand{\R}{\mathbb{R}}

\newcommand{\w}{\omega}

\gdef \RR{{\Bbb R}}

\makeatletter
\def\Ddots{\mathinner{\mkern1mu\raise\p@
\vbox{\kern7\p@\hbox{.}}\mkern2mu
\raise4\p@\hbox{.}\mkern2mu\raise7\p@\hbox{.}\mkern1mu}}
\makeatother

\begin{document}
\begin{frontmatter}

\title{Fixed-order strong H-infinity control of interconnected systems with time-delays}

\author[First]{Suat Gumussoy},
\author[First]{Wim Michiels}

\address[First]{Department of Computer Science, K. U. Leuven, \\
        Celestijnenlaan 200A, 3001, Heverlee, Belgium \\
        \mbox{(e-mails: suat.gumussoy@cs.kuleuven.be, wim.michiels@cs.kuleuven.be)}.}

\begin{abstract}
We design fixed-order strong H-infinity controllers for general time-delay systems. The designer chooses the controller order and may introduce constant time-delays in the controller. We represent the closed-loop system of the plant and the controller as delay differential algebraic equations (DDAEs). This representation deals with any interconnection of systems with time-delays without any elimination techniques. We present a numerical algorithm to compute the strong H-infinity norm for DDAEs which is robust to arbitrarily small delay perturbations, unlike the standard H-infinity norm. We optimize the strong H-infinity norm of the closed-loop system based on non-smooth, non-convex optimization methods using this algorithm and the computation of the gradient of the strong H-infinity norm with respect to the controller parameters. We tune the controller parameters and design H-infinity controllers with a prescribed order or structure.
\end{abstract}

\begin{keyword}
fixed-order controller design, strong h-infinity norm, time-delay, interconnected systems, delay differential algebraic equations, computational methods.
\end{keyword}

\end{frontmatter}

\section{Introduction}
In control applications, robust controllers are desired to achieve stability and performance requirements under model uncertainties and exogenous disturbances, \cite{zhou}. The design requirements are usually defined in terms of $\Hi$ norms of the closed-loop functions including the plant, the controller and weights for uncertainties and disturbances. There are robust control methods to design the optimal $\Hi$ controller for linear finite dimensional multi-input-multi-output (MIMO) systems based on Riccati equations and linear matrix inequalities (LMIs), see e.g.~\cite{DGKF, GahinetApkarian_HinfLMI} and the references therein. The order of the controller designed by these methods is typically larger or equal than the order of the plant. This is a restrictive condition for high-order plants, since low-order controllers are desired in a practical implementation. The design of fixed-order or low order $\Hi$ controller design can be translated into a non-smooth, non-convex optimization problem.  Recently fixed-order $\Hi$ controllers have been successfully designed for finite dimensional LTI MIMO plants using a direct optimization approach, \cite{suatHIFOO}. This approach allows the user to choose the controller order and tunes the parameters of the controller to minimize the $\Hi$ norm of the objective function. An extension to a class of retarded time-delay systems has been described in \cite{bfgbookchapter}.

We design a fixed-order $\Hi$ controller in a feedback connection with a time-delay system. The closed-loop system is a delay differential algebraic system and its state-space representation is written as
\begin{equation}\label{system}
\left\{\begin{array}{l}
E \dot x(t)= A_0 x(t)+\sum_{i=1}^m A_i x(t-\tau_i) +B w(t), \\
z=C x(t).
\end{array}\right.
\end{equation}
The time-delays $\tau_i$, $i=1,\ldots,m$ are positive real numbers and the capital letters are real-valued matrices with appropriate dimensions. The input $w$ and output $z$ are disturbances and signals to be minimized to achieve design requirements and some of system matrices include the controller parameters.

The system with the closed-loop equations (\ref{system}) represents all interesting cases of the feedback connection of a time-delay plant and a controller. The transformation of the closed-loop system to this form can be easily done by first augmenting the system equations of the plant and controller. As we shall see, this augmented system can subsequently be brought in the form (\ref{system}) by introducing slack variables to eliminate input/output delays and direct feedthrough terms in the closed-loop equations. Hence, the resulting system of the form (\ref{system}) is obtained directly without complicated elimination techniques, that may even not be possible in the presence of time-delays.

By interconnecting systems and controller high frequency paths could be created in control loops, which may lead to sensitivity problems with respect to the delays and delay perturbations. Therefore it is important to take the sensitivity explicitly into account in the design. As shown in \cite{TW578:10} the $\Hi$ norm of a DAE is not robust against small delay changes. This leads to the definition of the strong $\Hi$ norm, the smallest upper bound of the $\Hi$ norm that is insensitive to small delay changes, which are inevitable in any practical design due to small modeling errors. Several properties of the strong $\Hi$ norm are given in \cite{TW578:10}. In this paper, we present a level set algorithm for computing strong $\Hi$ norms, see~\cite{byers,boydbala,steinbuch} for the underlying idea behind level set methods. Since time-delay systems are inherently infinite-dimensional systems we adopt a predictor-corrector approach, where the prediction step involves a finite-dimensional approximation of the problem, and the correction step serves to remove the effect of the discretization error on the numerical result.

The  numerical algorithm for the norm computation is subsequently applied to the design of fixed-order $\Hi$ controllers by a direct optimization approach. In the context of control of  LTI systems it is well known that  $\Hi$ norms are in general  non-convex  functions of the controller parameters which arise as elements of the closed-loop system matrices. They are typically even not everywhere smooth, although they are differentiable almost everywhere, \cite{suatHIFOO}. These properties carry over to the case of strong $\Hi$ norms of DDAEs under consideration. Therefore, special optimization methods for non-smooth, non-convex problems are required. We will use a combination of BFGS, whose favorable properties in the context of non-smooth problems have been reported in \cite{overtonbfgs}, bundle and gradient sampling methods, as implemented in the MATLAB code HANSO\footnote{Hybrid Algorithm for non-smooth Optimization, see~\cite{overtonhanso}}. The overall algorithm only requires the evaluation of the objective function, i.e.,~the strong $\Hi$ norm, as well as its derivatives with respect to the controller parameters whenever it is differentiable. The computation of the derivatives is also discussed in the paper.

The presented method is frequency domain based and builds on the eigenvalue based framework developed in~\cite{bookwim}.  Time-domain methods for the $\Hi$ control of DDAEs have been described in \cite{fridman} and the references therein, based on the construction of Lyapunov-Krasovskii functionals.

\section{Preliminaries}
Let $\mathrm{rank}(E)=n-v$, with $v\leq n$, and let the columns
of matrix $U\in\RR^{n\times v}$, respectively $V\in\RR^{n\times v}$, be a (minimal) basis for
the left, respectively right nullspace, that is,
\begin{equation}\label{nullspace}
U^T E=0,\ \ E V=0.
\end{equation}
Throughout the paper we make the following assumptions.
\begin{assum} \label{assumption}
The matrix $U^T A_0 V$ is nonsingular.
\end{assum}

\begin{assum} \label{assumption_sstab}
The zero solution of system (\ref{system}), with $w\equiv0$, is strongly exponentially stable.
\end{assum} Assumption \ref{assumption} prevents that the equations (\ref{system}) are of advanced type  and, hence, non-causal. Assumption \ref{assumption_sstab} guarantees that the asymptotic stability of the null solution is robust against small delay perturbations, \cite{Hale:2002:STRONG}.

Finally we use the following notations. Set of nonnegative and strictly positive real numbers are $\R^+,\R_0^+$. The $m$-tuple time-delays $(\tau_1,\ldots,\tau_m)$ are denoted as $\vec\tau$. The open ball of radius $\epsilon\in\R^+$ centered at $\vec\tau\in$ $(\R^+)^m$ is shown as $\mathcal{B}(\vec \tau,\epsilon):=\{\vec\theta\in(\R)^m : \|\vec\theta-\vec \tau\|<\epsilon\}$. The i$^\mathrm{th}$ singular value of a matrix $A$ is denoted by $\sigma_i(A)$.

\section{Motivating example} \label{sec:motex}
We give the following motivating example to illustrate the generality of the system description (\ref{system}).
\begin{exmp} \label{motex}
Consider the feedback interconnection of the system
\[
\left\{\begin{array}{rll}
\frac{d}{dt}\left(x(t)+Hx(t-\tau_3)\right) &=& A x(t)+B_0 w(t) \\
&& \hspace{1cm} +B_1 w(t-\tau_1)+B_2u(t),\\
z(t)&=& C_z x(t)+D_z w(t), \\
y(t)&=& C_y x(t)+D_u u(t),
\end{array}\right.
\]
and the controller
\[
u(t)=K y(t-\tau_2).
\]
\end{exmp}

The feedback connection has a delay, $\tau_2\neq 0$, therefore we can not eliminate the output and controller equation. The system is a neutral time-delay system with a direct feedthrough term from $w$ to $z$, i.e., $D_z$ and an input delay $\tau_1$. By defining a new state $X$ including $u$ and $y$ of the system variables with $\gamma_x$ and $\gamma_w$  slack variables, we can easily transform the feedback connection into the form (\ref{system}).

If we let $X=[\gamma_x^T\ x^T\ u^T\ y^T\ \gamma_w^T]^T$, we can describe the system by the equations
{\small
\[\left\{
\begin{array}{l}
\underbrace{
\left[
\begin{array}{cc}
I & 0_{1\times5} \\
0_{5\times 1} &0_{5\times5}
\end{array}
\right]
}_{E}
\dot X(t)=
\underbrace{
\left[
\begin{array}{cccccc}
0 & A & B_2 & 0 & 0 & B_0\\
0 & 0 & -I & 0 & 0 & 0\\
0 & C_y & D_y & 0 & -I & 0\\
-I & I & 0 & 0 & 0 & 0\\
-I & I & 0 & 0 & 0 & 0\\
0 & 0 & 0 & 0 & 0 & -I
\end{array}
\right]
}_{A_0}
X(t) \\
\hspace{.9cm}+
\underbrace{
\left[
\begin{array}{cc}
0_{1\times5} & B_1 \\
0_{5\times5} & 0
\end{array}
\right]
}_{A_1} X(t-\tau_1)
+
\underbrace{
\left[
\begin{array}{ccc}
0_{1\times4} & 0 & 0 \\
0_{1\times4} & K & 0 \\
0_{4\times4} & 0_{4\times1} & 0_{4\times1}
\end{array}
\right]
}_{A_2}
X(t-\tau_2) \\
\hspace{2.55cm}+
\underbrace{
\left[\begin{array}{ccc}
0_{3\times1} & 0_{3\times1} & 0_{3\times3} \\
0 & H & 0_{1\times3} \\
0_{2\times1} & 0_{2\times1} & 0_{2\times3}
\end{array}
\right]}_{A_3}
X(t-\tau_3)
+
\underbrace{
\left[\begin{array}{c}
0 \\
0 \\
I
\end{array}
\right]}_{B}
w(t) \\
z=
\underbrace{\left[
\begin{array}{cccc}
0 & C_z & 0_{1\times 3} & D_z
\end{array}
\right]}_{C} X(t)
\end{array}
\right.
\]
}where each element of the above matrices is a block matrix of appropriate dimensions.

Using the technique illustrated with the above example, a broad class of interconnected
systems with delays can be brought in the form (\ref{system}), where the external
inputs $w$ and outputs $z$ stem from the performance specifications expressed in terms of
appropriately defined transfer functions. The price to pay for the generality of the framework is the increase of the dimension of the system, $n$, which affects the efficiency of the numerical methods. However, this is a minor problem in most applications because the delay difference equations or algebraic constraints are related to inputs and outputs, and
the number of inputs and outputs is usually much smaller than the number of state variables.

\section{Strong $\Hi$ Norm and Its Computation}

\subsection{Definitions and properties}

We write the transfer function of the system (\ref{system}) as
\begin{equation} \label{T}
T(\lambda):=C\left(\lambda E-A_0-\sum_{i=1}^m A_i e^{-\lambda \tau_i}\right)^{-1}B
\end{equation}
and define the {\it asymptotic} transfer function of the system (\ref{system}) as
\begin{equation} \label{Ta}
T_a(\lambda):=-C V \left(U^T A_0 V +\sum_{i=1}^m U^T A_i V e^{-\lambda\tau_i} \right)^{-1} U^TB.
\end{equation}

The terminology stems from the fact that the transfer function $T$ and the asymptotic transfer function $T_a$ converge to each other for high frequencies, see \cite{TW578:10}.

In \cite{TW578:10}, it is shown that the function
\begin{equation}\label{defTdel}
\vec\tau\in(\RR_{0}^+)^{m}\mapsto \|T(j\w,\vec\tau)\|_\infty,
\end{equation}
is, in general, not continuous, which is inherited from the behavior of the asymptotic transfer function, $T_a$, more precisely the function
\begin{equation}\label{defTadel}
\vec\tau\in(\RR_{0}^+)^{m}\mapsto \|T_a(j\w,\vec\tau)\|_\infty.
\end{equation}

Since small modeling errors and uncertainty are inevitable in a practical design, we are interested in the smallest upper bound for the $\Hi$ norm which is insensitive to small delay perturbations. A formal definition of the \emph{strong $\Hi$ norm} is as follows.

\begin{defn} \label{def:shinfTa}
Let $G(\lambda;\vec \tau)$ be the transfer function of a strongly stable system. The strong $\mathcal{H}_{\infty}$ norm of $G$, $\interleave {G}(j\w,\vec \tau)\interleave_\infty$, is defined as
\begin{multline} \nonumber
\interleave G(j\w,\vec \tau)\interleave_\infty:=\lim_{\epsilon\rightarrow 0+}
\sup \{\|G(j\w,\vec \tau_\epsilon)\|_\infty: \\ \vec \tau_\epsilon\in\mathcal{B}(\vec \tau,\epsilon) \cap (\RR^+)^{m} \}.
\end{multline}
\end{defn}

Several properties of the strong $\Hi$ norm of $T$ and $T_a$ are listed below.
\begin{prop}\label{prop:Tasinfprop}
The following assertions hold \cite{TW578:10}.
\begin{enumerate}
\item For every $\vec\tau\in(\RR_0^+)^m$, we have
\begin{equation} \label{Tasweep}
\interleave T_a(j\w,\vec \tau)\interleave_\infty=\max_{\vec\theta\in [0,\
2\pi]^m} \sigma_{1} \left( \mathbb{T}_a(\vec \theta) \right),
\end{equation}
where
\begin{equation} \label{Ta_theta}
\mathbb{T}_a(\vec \theta):=C V \left(-U^T A_0 V -\sum_{i=1}^m U^T A_i V e^{-j\theta_i} \right)^{-1} U^TB.
\end{equation}
\item The function
\begin{equation}\label{shinfnorm2}
    \vec \tau\in(\RR^+_0)^m\mapsto \interleave T(j\w,\vec \tau)\interleave_\infty
\end{equation}
is continuous.
\item  The strong $\Hi$ norm of the transfer function $T$ satisfies
\begin{equation} \label{shinfnorm}
    \interleave T(j\w,\vec \tau)\interleave_\infty=\max\left(\|T(j\w,\vec\tau)\|_{\infty}, \interleave T_a(j\w,\vec \tau)\interleave_\infty \right).
\end{equation}
\item Let $\xi>\interleave T_a(j\w,\vec \tau)\interleave_\infty$ hold.
Then there exist real numbers $\epsilon>0,\ \Omega>0$ and an integer $N$ such that for any $\vec r\in\mathcal{B}(\vec\tau,\epsilon)\cap(\mathbb{R}^+)^m$, the number of frequencies $\w^{(i)}$ such that
\begin{equation}
\sigma_k\left(T(j\w^{(i)},\vec r)\right)=\xi,
\end{equation}
for some $k\in\{1,\ldots,n\}$, is smaller then $N$, and, moreover, $|\w^{(i)}|<\Omega$.
\end{enumerate}
\end{prop}

The strong $\Hi$ norm of the transfer function $T$ can be computed by (\ref{shinfnorm}) depending on the computation of the $\Hi$ norm of $T$ and the strong $\Hi$ norm of $T_a$. Therefore, in~\S\ref{sec:hinfnorm_Ta}, we first give a numerical method for the strong $\Hi$ norm computation of the asymptotic transfer function $T_a$ based on the computational formula (\ref{Tasweep}) of Proposition \ref{prop:Tasinfprop}. Next, we present the algorithm for computing the strong $\Hi$ norm of $T$ in~\S\ref{sec:shinfnorm_T}.

\subsection{Strong $\Hi$ norm of the asymptotic transfer function} \label{sec:hinfnorm_Ta}

The computation of $\interleave T_a(j\w,\vec \tau)\interleave_\infty$ is based on expression (\ref{Tasweep}).  We obtain an approximation by restricting $\vec \theta$ in (\ref{Tasweep}) to a grid,
\begin{equation} \label{Taapprox}
\interleave T_a(j\w,\vec \tau)\interleave_\infty\approx\max_{\vec\theta\in\Omega_h} \sigma_{1} \left(\mathbb{T}_a(\vec \theta)\right),
\end{equation} where $\Omega_h$ is a m-dimensional grids over the hypercube $[0,\ 2\pi]^m$ and $\mathbb{T}_a(\vec \theta)$ is defined by (\ref{Ta_theta}).
If a high accuracy  is required, then the approximate results may be corrected to the full precision by solving the nonlinear equations
\begin{equation} \label{eq:Tacorrection}
\left\{\begin{array}{l}
\left[\begin{array}{cc}
\mathbb{A}_{22}(\vec \theta) & -\xi^{-1}U^TBB^TU\\
\xi^{-1} V^TC^TCV & -\mathbb{A}_{22}^{*}(\vec \theta)
\end{array}\right]
\left[\begin{array}{c}
u_a \\
v_a
\end{array}\right]
=0, \\
n(u_a,v_a)=0, \\
\Re ( e^{-j\theta_i}(v_a^* U^TA_iV u_a) )=0,\ i=1,\ldots,m,
\end{array}\right.
\end{equation}
where
\[
 \mathbb{A}_{22}(\vec \theta)=-U^T A_0 V -\sum_{i=1}^m U^T A_i V e^{-j\theta_i}
 \]
and $n(u_a,v_a)=0$ is a normalization constraint. The first equation in (\ref{eq:Tacorrection}) implies that  $\xi$ is a singular value of $\mathbb{T}_a(\vec\theta)$. The last equation of (\ref{eq:Tacorrection}) expresses that the derivatives of the singular value $\xi$ with respect to the elements of $\vec \theta$ are zero.
In our implementation we solve (\ref{eq:Tacorrection}) using the Gauss-Newton method, which exhibits quadratic convergence because the (overdetermined) equations have an exact solution.

In most practical  problems, the number of delays to be considered in $\mathbb{A}_{22}(\vec \theta)$ is much smaller than the number of system delays, $m$, because most of the terms are zero. Note that in a control application a nonzero term corresponds to a high frequency feedthrough over the control loop.

\subsection{Algorithm} \label{sec:shinfnorm_T}
From (\ref{shinfnorm}) the following implication can be derived.
\begin{multline}
\nonumber \interleave T(j\w,\vec \tau)\interleave_\infty> \interleave T_a(j\w,\vec \tau)\interleave_\infty \Rightarrow \\
\interleave T(j\w,\vec \tau)\interleave_\infty=\|T(j\w,\vec \tau)\|_\infty.
\end{multline}
Moreover, we know from Statement (4) of Proposition \ref{prop:Tasinfprop} that, given a level
\begin{equation}\label{over}
\xi>\interleave T_a(j\w,\vec \tau)\interleave_\infty,
\end{equation}
there are only \emph{finitely} many frequencies $\omega$ for which for a singular value of $T(j\omega,\vec\tau)$ is equal to $\xi$. These properties allow a slight adaptation of the level set algorithm for $\Hi$ computations of retarded time-delay systems as described in~\cite{wimsimax}, whenever one restricts to the situation where (\ref{over}) holds. The latter is possible by a preliminary computation of the strong $\Hi$ norm of $T_a$, as outlined in \S\ref{sec:hinfnorm_Ta}.

The level set method is based on a predictor-corrector approach. In the prediction (approximation) step the infinite-dimensional problem is discretized allowing to apply methods for LTI systems. In particular, the time-delay system (\ref{system}) can be approximated by a finite-dimensional system using a spectral method, as in \cite{jorisIJC}. The finite-dimensional system is described as \begin{eqnarray}\label{finsystem}
{\bf E_N}\dot z(t)&=&{\bf A_N} z(t)+{\bf B_N} u(t),\ z(t)\in\R^{(N+1)n\times 1} \\
y(t)&=&{\bf C_N}z(t)
\end{eqnarray} where $N$ is a positive integer for the number of discretization points in the interval $[-\tau_{\max},\ 0]$. The transfer function of (\ref{finsystem}) is given by
\begin{equation} \label{eq:TN}
T_N(\lambda):={\bf C_N}(\lambda {\bf E_N}-{\bf A_N})^{-1}{\bf B_N}.
\end{equation}
Further details on the transformation to and the infinite-dimensional system and the discretization into a finite-dimensional system are given in \cite{wimsimax}.

In the correction step the effect of the approximation on the computed $\Hi$ norm is removed. The following algorithm computes the strong $\Hi$ norm within the tolerance, tol.

\subsubsection*{Algorithm}
${}$\\ \\
\underline{\emph{Prediction step:}}
    \begin{enumerate}
    \item calculate the first level, $\xi_l=\interleave T_a(j\w,\vec \tau)\interleave_\infty$,
    \item repeat until break
        \begin{enumerate}
        \item set $\xi:= \xi_l(1 + 2 \mathrm{tol})$
        \item compute all $\w^{(i)}\in\R$ satisfying $\sigma_k \left(T_N(j\w^{(i)})\right)=\xi$. By \cite[Proposition 12]{genin}, this can be done by computing generalized eigenvalues of the pencil
            {\small
            \begin{equation}\label{pencil}
                \lambda
                \left[
                    \begin{array}{cc}
                        {\bf E_N}& 0\\
                        0 & {\bf E_N}^T
                    \end{array}
                \right]-
                \left[
                    \begin{array}{cc}
                        {\bf A_N}& \xi^{-1}{\bf B_N B_N}^T\\
                        -\xi^{-1}{\bf C_N}^T{\bf C_N} & -{\bf A_N}^T
                    \end{array}
                \right],
            \end{equation}}
            whose imaginary axis eigenvalues are given by $\lambda=j\w^{(i)}$.
        \item \textbf{\em if} no generalized eigenvalues $j\w^{(i)}$ of (\ref{pencil}) exist,
            \textbf{\em then}
              \begin{enumerate}
                  \item[] \textbf{\em if} $\xi_l=\interleave T_a(j\w,\vec \tau)\interleave_\infty$, \textbf{\em then} \\
                      \hspace*{.5cm}set
                       $
                       \interleave T(j\w,\vec\tau)\interleave_\infty=
                       \interleave T_a(j\w,\vec\tau)\interleave_\infty
                       $\\
                       \hspace*{.5cm}quit
                  \item[] \textbf{\em else} \\
                  \hspace*{.5cm} compute $\w^{(i)}\in\R$ satisfying \\
                  \hspace*{3.5cm}  $\sigma_k \left(T_N(j\w^{(i)})\right)=\xi_l$,\\
                  \hspace*{.5cm} set $\tilde \xi=(\xi+\xi_l)/2$,  $\tilde{\w}^{(i)}=\w^{(i)}, i=1,2,\ldots$ \\
                  \hspace*{.5cm} break, go to correction step.\\
                  \textbf{\em endif}
              \end{enumerate}
              \textbf{\em else}
              \begin{enumerate}
                  \item[] calculate $\mu^{(i)}:=\sqrt{\w^{(i)}\w^{(i+1)}}$, $i=1,2,\ldots$
                  \item[] set
                  \begin{multline*}
                  \xi_l:=\max_i \max \left(\sigma_{1}\left( T_N(j\mu^{(i)})\right), \right.\\
					\left. \interleave T_a(j\w,\vec \tau)\interleave_\infty \right).
                  \end{multline*}
              \end{enumerate}
              \textbf{\em endif}
        \end{enumerate}
${}$\\
\underline{\emph{Correction step:}} \\
%
\begin{enumerate}
\item Solve the nonlinear equations
\begin{equation} \label{eq:Tcorrection}
\left\{\begin{array}{l}H(j\omega,\xi)
\left[\begin{array}{c}
u \\
v
\end{array}\right]
=0, \\
n(u,v)=0, \\
\Im \{v^* (E+\sum_{i=1}^m A_i\tau_ie^{-j\w\tau_i})u \}=0,
\end{array}\right.
\end{equation}
where
{\scriptsize
\begin{multline*}
H(j\omega,\xi)=\\
\left[\begin{array}{cc}
j\omega E-A_0-\sum_{i=1}^m A_i e^{-j\omega\tau_i} & -\xi^{-1}BB^T\\
\xi^{-1}C^TC & (j\omega E+A_0+\sum_{i=1}^m A_i e^{j\omega\tau_i})^T
\end{array}\right]
\end{multline*}} and $n(u,v)=0$ is a normalizing condition, with the starting values
\begin{multline*}
\omega=\tilde\omega^{(i)},\ \ \xi=\tilde\xi\; \textrm{and} \\
\left[\begin{array}{c}u\\
v\end{array}\right]=
\arg\min{\|H(j\tilde\omega^{(i)},\tilde\xi)v\|}/{\|v\|};
 \end{multline*}
denote the solutions with $(\hat u^{(i)},\hat
 v^{(i)},\hat\omega^{(i)},\hat\xi^{(i)})$, for $i=1,2,...,$
\item set
$\interleave T(j\omega)\interleave_\infty:=\max_{1\leq i\leq
p}\hat \xi^{(i)}$
\end{enumerate}
%
\end{enumerate}

\medskip

\begin{figure}[!h]
    \begin{minipage}[t]{0.45\textwidth}
        \vspace{0pt}
        \begin{center}
        \includegraphics[width=0.8\linewidth]{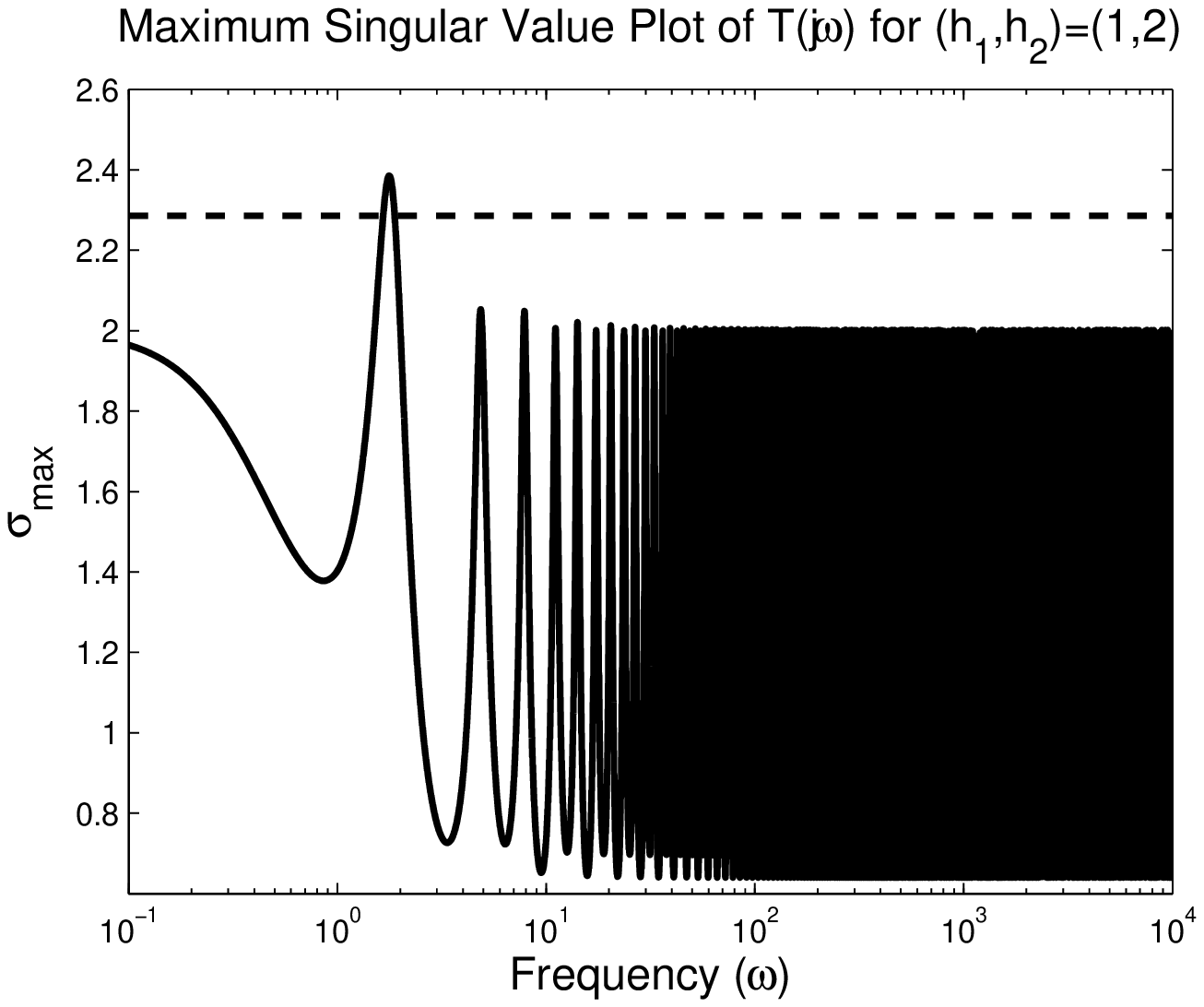}
        \caption{\label{fig:svd100} The maximum singular value plot of $T(j\w)$ for $(\tau_1,\tau_2)=(1,2)$ as a function of $\omega$.}
        \end{center}
    \end{minipage}
\hfill
    \begin{minipage}[t]{0.45\textwidth}
        \vspace{0pt}\raggedright
        \begin{center}
        \includegraphics[width=0.8\linewidth]{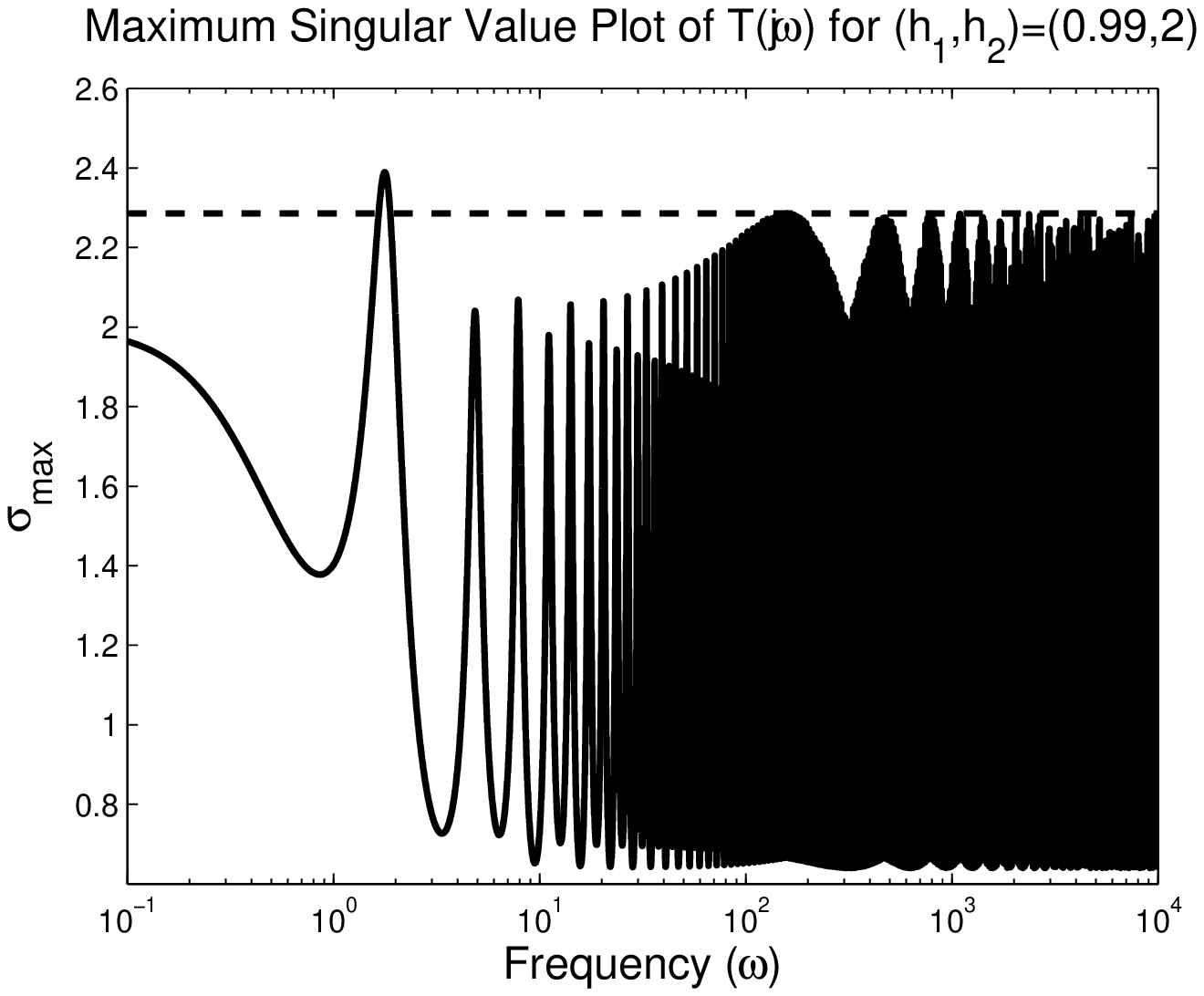}
        \caption{\label{fig:svd099} The maximum singular value plot of $T(j\w)$ for $(\tau_1,\tau_2)=(0.99,2)$ as a function of $\omega$.}
        \end{center}
   \end{minipage}
\end{figure}

The computational cost of Algorithm in the prediction step is dominated by the computation of the generalized  eigenvalues of the pencil matrices with dimensions \mbox{$(N + 1)2n$}. Mathematically, equations (\ref{eq:Tcorrection}) characterize extrema in the singular value curves (see~\cite{wimsimax}), and hence, they can be used to correct peak values. Note that the correction step is only performed if
\[
\interleave T(j\omega,\vec\tau)\interleave_{\infty}>\interleave T_a(j\omega,\vec\tau)\interleave_{\infty}.
\] In our implementation we solve equations (\ref{eq:Tcorrection}) in least squares sense using
the Gauss Newton algorithm, which can be shown to be quadratically converging in
the case under consideration where the residual in the desired solution is zero.

For details on the choice of the number of discretization points, $N$, and the tolerance, tol, we refer to~\cite{wimsimax}.

\begin{exmp}
Figures \ref{fig:svd100} and \ref{fig:svd099} show singular value plots of the transfer function of $(\tau_1,\tau_2)=(1,2)$ and $(\tau_1,\tau_2)=(0.99,2)$ for
\[
T(\lambda,\vec\tau):=\frac{\lambda+2}{\lambda(1-1/16e^{-\lambda\tau_1}+1/2e^{-\lambda\tau_2})+1}.
\] The strong $\Hi$ norm of the asymptotic transfer function $T_a$ is shown as dashed lines. We use this value as an initial level in the Algorithm. This example also illustrates that the $\Hi$ norm of the asymptotic transfer function of a time-delay system may be sensitive to small delay changes as shown in Figures \ref{fig:svd100} and \ref{fig:svd099}.
\end{exmp}

\smallskip

\section{Fixed-order $\Hi$ controller design} \label{sec:design}

The closed-loop system is described as
\[
\begin{array}{l}
E \dot x(t)=A_0(p) x(t)+\sum_{i=1}^m A_i(p) x(t-\tau_i)+Bw(t) \\
z(t)=Cx(t)
\end{array}
\] where the vector $p$ contains all the parameters in the controller matrices. We design fixed-order $\Hi$ controllers by minimizing the strong $\Hi$ norm of the closed-loop transfer function $T$ as a function of $p$. This is a non-convex problem and the objective function (the strong $\Hi$ norm) with respect to optimization parameters (controller parameters) is a non-smooth function but its differentiable almost everywhere. Given these properties, we use the non-smooth, non-convex optimization method proposed in \cite{suatHIFOO} and implemented as a MATLAB function HANSO in \cite{overtonhanso}. The optimization algorithm searches for the local minimizer of the objective function in three steps: a quasi-Newton algorithm (in particular, BFGS) to
approximate a local minimizer; a local bundle method to verify local optimality for the best point found by BFGS; if this does not succeed, gradient sampling to refine the approximation of the local minimizer, \cite{BurkeTAC06}. The optimization algorithm requires the evaluation of the objective function and its gradients with respect to the optimization parameters, whenever it is differentiable. These are described now.

The strong $\Hi$ norm of the transfer function $T$ other corresponding parameters are computed by Algorithm \ref{sec:shinfnorm_T}. The derivatives of the norm with respect to controller parameters exist whenever there are unique time-delay values $\vec{\hat{\theta}}$ or a frequency $\hat\omega$ such that
\[
\interleave T(j\w,\vec\tau)\interleave_\infty=\hat\xi=\left\{\begin{array}{ll}
\sigma_{1}(\mathbb{T}_a(\vec{\hat \theta})), & \mathrm{if}\ \hat \xi=\interleave T_a(j\w,\vec\tau)\interleave_\infty, \\
\sigma_{1}(\mathbb{T}(j\hat\w)), & \mathrm{if}\ \hat \xi>\interleave T_a(j\w,\vec\tau)\interleave_\infty
\end{array}\right.
\] holds and, in addition, the largest singular value $\hat\xi$ has multiplicity one. We compute the derivative of the strong $\Hi$ norm of $T$ with respect to the optimization parameter $p_i$ in the controller matrices as
\[
\frac{\partial \xi}{\partial p_i}=\left\{
\begin{array}{l}
-2\xi^2\left.\frac{\Re\left(v_a^*\frac{\partial \mathbb{A}_{22}(\vec \theta)}{\partial p_i} u_a\right)}
{v_a^*U^TBB^TUv_a+u_a^*V^TC^TCVu_a}\right|_{(\xi,\vec \theta)=(\hat{\xi},\vec{\hat{\theta}})}  \\
\hspace*{3.5cm}\mathrm{if}\ \ \hat{\xi}=\interleave T_a(j\w,\vec \tau)\interleave_\infty,\\
-2\xi^2\left.\frac{\Re\left(v^*\frac{\partial A(j\w)}{\partial p_i} u\right)}
{v^*BB^Tv+u^*C^TCu}\right|_{(\xi,\w)=(\hat{\xi},\hat{\w})} \\
\hspace*{3.5cm} \mathrm{if}\ \hat \xi>\interleave T_a(j\w,\vec\tau)\interleave_\infty
\end{array}\right.
\] where given $\xi=\hat \xi$, $u_a,v_a$ and $u,v$ are vectors in (\ref{eq:Tacorrection}) and (\ref{eq:Tcorrection}) for $\vec\theta=\vec{\hat{\theta}}$ and $\w=\hat\w$ respectively. For detailed derivation on derivative calculations, see \cite{thesismarc,bfgbookchapter}.

We note that our approach allows constant entries in the controller matrices. Hence, we can impose a structure on the controller, e.g., a PID controller. Although we illustrated our method for a dynamic controller, it can be applied to more general controller structures including time-delays in the controller states or inputs.

\section{Examples} \label{sec:ex}

Consider the feedback interconnection of the system
\[
\begin{array}{rll}
\dot{x}(t) &=& \left(
                  \begin{array}{cc}
                    2 & 1 \\
                    0 & -1 \\
                  \end{array}
                \right)
 x(t)
 +
 \left(
    \begin{array}{cc}
      -1 & 0 \\
      -1 & 1 \\
    \end{array}
 \right)
 x(t-h)+ \\
 && \hspace{3cm} \left(
    \begin{array}{c}
      -0.5 \\
      1 \\
    \end{array}
 \right)
 w(t)
 + \left(
    \begin{array}{c}
      3 \\
      1 \\
    \end{array}
 \right)
 u(t) \\
z(t)&=&
\left(
  \begin{array}{cc}
    1 & -0.5 \\
    0 & 0 \\
  \end{array}
\right)
x(t)+
 \left(
    \begin{array}{c}
      0 \\
      1 \\
    \end{array}
 \right)
u(t), \\
y(t)&=& x(t),
\end{array}
\]
and the controller
\[
u(t)=K y(t).
\]

In \cite{fridman98}, a static order controller is designed with $\Hi$ performance $0.4436$ for the given delays in Table~\ref{table:ex}. We designed static order controllers using our approach and give their closed-loop $\Hi$ performances for various delays in Table~\ref{table:ex}. We will present our extensive benchmark results in the full version of the paper.

\begin{table}
\begin{center}
\begin{tabular}{ccc}
  \hline
  \hline
  h & $\xi$ & K \\
  \hline
  $0.1$ & $0.4005$ & $[-17.8065,\ 9.5915]$\vspace{.5mm}\\
  $0.2$ & $0.3981$ & $[-7.1854,\ 3.7727]$\vspace{.5mm}\\
  $0.3$ & $0.3995$ & $[-4.3068,\ 2.0695]$\vspace{.5mm}\\
  $0.4$ & $0.4041$ & $[-3.7321,\ 1.6556]$\vspace{.5mm}\\
  $0.5$ & $0.4101$ & $[-3.5878,\ 1.5017]$\vspace{.5mm}\\
  $0.6$ & $0.4158$ & $[-3.4104,\ 1.3563]$\vspace{.5mm}\\
  $0.7$ & $0.4206$ & $[-3.2772,\ 1.2514]$\vspace{.5mm}\\
  $0.8$ & $0.3953$ & $[0.8892,\ -0.9308]$\vspace{.5mm}\\
  $0.9$ & $0.3953$ & $[0.0518,\ -0.4074]$\vspace{.5mm}\\
  $1.0$ & $0.3953$ & $[0.1942,\ -0.4964]$\vspace{.5mm}\\
  \hline
  \hline
\end{tabular}
\end{center}
\caption{The achieved $\Hi$ performances $\xi$ by static order controllers} \label{table:ex}
\end{table}

\section{Concluding Remarks} \label{sec:conc}
We showed that a very broad class of interconnected systems can be brought in the standard form (\ref{system}) in a systematic way. Input/output delays and direct feedthrough terms can be dealt with by introducing slack variables. An additional advantage in the context of control design is the linearity of the closed loop matrices w.r.t. the controller parameters.

We presented a predictor-corrector algorithm for the strong $\Hi$ norm computation of DDAEs. Based on the numerical algorithm for the strong $\Hi$ norm and its gradient computation with respect to controller parameters, we applied non-smooth, non-convex optimization methods for designing controllers with a fixed-order or structure.

\section*{Acknowledgements}
This work has been supported by the Programme of Interuniversity Attraction Poles of the Belgian Federal Science Policy Office (IAP P6- DYSCO), by OPTEC, the Optimization in Engineering Center of the K.U.Leuven, by the project STRT1-09/33 of the K.U.Leuven Research Council and the project G.0712.11N of
the Research Foundation - Flanders (FWO).

\bibliography{referentielijst,otherref}            

\end{document}